\documentclass[pdflatex,sn-mathphys-num]{sn-jnl}

\usepackage{graphicx}
\usepackage{multirow}
\usepackage{amsmath,amssymb,amsfonts}
\usepackage{amsthm}
\usepackage{mathrsfs}
\usepackage[title]{appendix}
\usepackage{xcolor}
\usepackage{textcomp}
\usepackage{manyfoot}
\usepackage{booktabs}
\usepackage{algorithm}
\usepackage{algorithmicx}
\usepackage{algpseudocode}
\usepackage{listings}
\usepackage{hyperref}

\usepackage{tikz}
\usetikzlibrary{trees, calc, decorations.pathreplacing}

\usepackage{eucal}          
\usepackage{amscd}          
\usepackage{latexsym}
\usepackage{mathtools}      
\usepackage{dsfont}

\usepackage[all]{xy}

\usepackage{tcolorbox}

\usepackage{float}          
\usepackage{multicol}

\usepackage{subcaption}

\usepackage{epigraph}

\theoremstyle{thmstyleone}
\newtheorem{thm}{Theorem}[section]

\newtheorem{prop}[thm]{Proposition}

\theoremstyle{thmstyletwo}

\theoremstyle{thmstylethree}
\newtheorem{definition}[thm]{Definition}

\newtheorem*{theorem*}{Theorem}
\newtheorem*{corollary*}{Corollary}

\newtheorem*{Satz*}{Satz}

\newtheorem*{Proposal*}{Proposal}

\begin{document}

\title[]{Dynamic centrality of headwater sources in river networks: a stochastic approach via ultrametric Laplacians}

\author*[1]{\fnm{Ángel Alfredo} \sur{Morán Ledezma}}\email{angel.ledezma@kit.edu}

\affil*[1]{\orgdiv{Geodetic Institute}, \orgname{Karlsruhe Institute of Technology},
  \orgaddress{\street{Englerstr. 7}, \city{Karlsruhe}, \postcode{76131},
  \state{Karlsruhe}, \country{Germany}}}

\abstract{\small\setlength{\baselineskip}{10pt}
River networks are hierarchical transport systems in which 
the timing and position of headwater confluences govern 
hydrologic response, solute transport, and ecological 
connectivity. Despite the recognized importance of headwater 
sources in structuring downstream processes, no mathematically 
grounded centrality index exists that captures their dynamic 
role in the transport hierarchy. We apply the dynamic 
centrality index $C_{\mathrm{CTMC}}$ \cite{MoranLedezma2026}, 
originally introduced in the context of phylogenetic trees, 
to the problem of headwater centrality in river networks via 
the dynamic tree representation of \cite{Zaliapin2010}. 
Through a topological analysis of the ultrametric structure 
induced by the dynamic tree, we show that high-centrality 
headwaters are the tributaries that most efficiently transmit 
water into the rest of the network, in the sense that their 
flows merge earliest and most broadly with surrounding 
sources as transport proceeds downstream. The index admits 
a fully explicit closed-form expression computable in $O(n)$ 
time from the tree structure alone, without simulation. 
Comparing $C_{\mathrm{CTMC}}$ rankings against the number 
of downstream junctions reached during transport, a 
direct measure of hydrological influence, on a dataset 
of 49 natural river basins across the United States, we 
find that top-ranked headwaters consistently reach a 
disproportionately large number of junctions across all 
transport times. This indicates that high-centrality 
headwaters are not merely early contributors but 
consistently influential throughout the entire transport 
process. These results suggest that ultrametric spectral 
analysis provides an interpretable and scalable framework 
for identifying hydrologically influential headwaters, 
with potential applications in ecological monitoring and 
watershed management.}


\maketitle

\section{Introduction}
River networks are archetypal examples of bifurcating transport systems, where their topology, is often described by hierarchical relationships. Their geometry and topology  has been recognized to be a key component on the timing, quality and magnitude of water, sediment, solutes, and organisms delivered to downstream locations.  \cite{Horton1945,Strahler1957,Shreve1966,Tokunaga1978,RodriguezIturbeRinaldo1997}. Beautiful geometric ideas appear in the morphology of the drainage networks; as established in the classic literature there exists quantitative scaling laws for stream numbers, lengths and areas  and the prevalence of self-similar patterns \cite{Horton1945,Strahler1957,Shreve1966,Tokunaga1978}.  These geometric regularities are reflected, for example, in the fractal and scaling properties of channel networks, giving rise to theories that relate the form of a basin to hydrologic function. \cite{Tarboton1988,Rigon1996,RodriguezIturbeRinaldo1997,GuptaWaymire1983}.

Many studies have linked network topology to hydrologic response.  Classical works have emphasized that hydrologic response emerges from the organization of flow paths and travel times in the channel network \cite{GuptaWaymire1983,RodriguezIturbeRinaldo1997}. Work on self-similar and fractal descriptions of river basins further shows that network geometry constrains the distribution of path lengths and confluence positions \cite{RodriguezIturbeRinaldo1997,Tarboton1988,Rigon1996,Tokunaga1978}, providing a natural link between network structure and the timing of downstream integration \cite{GuptaWaymire1983,Tarboton1988}. 

River networks structure not only influence the hydrologic transport but also geo- morphic, bio-chemical, ecological, and riparian processes. Their hierarchical, branching geometry governs the routing and mixing of water and sediment; the delivery and stor- age of wood and particulate material; the retention, transformation, and downstream export of nutrients; the spatial organization of riparian vegetation; the movement of aquatic organisms; and the architecture of food webs and biodiversity patterns. This processes occur thorough a directed tree network, hence the position of sources and timing of their connectivity produce different influences in these mechanisms, the dynamic and topological characteristic of the junctions have great influence in these processes.    \cite{Sklar2006,
Benda2004a,
Lowe2006,
Muneepeerakul2006,
PowerDietrich2002,
Rice2006,
BendaDunne1997,Fritz2018}

At the same time, a growing body of work emphasizes that individual tributary junctions can exert disproportionate control on physical habitat, transport, and ecosystem processes.  The ``network dynamics hypothesis'' of \cite{Benda2004a} argues that the river networks structure riverine habitats through a combination of stochastic watershed disturbances, local geomorphic responses at confluences, and the downstream propagation of these signals.  

In this view, the timing and position of confluences, particularly low-order or ``early'' junctions close to headwater sources, strongly influence the spatial pattern of flow resistance, bed morphology, and habitat heterogeneity emerging downstream \citep{Benda2004a,Rice2017}.

There is extensive evidence that headwater streams and riparian wetlands are physically and chemically connected to downstream waters. Structurally, this connectivity arises from the continuous channel network and the adjoining floodplain, which together make these systems physically contiguous. The degree to which water, nutrients, sediment, and contaminants are actually transferred downstream varies in space and time, depending on several factors like the distance to downstream waters. \citep{Fritz2018}. 

Headwater streams play a particularly important role in structuring biodiversity.  Although individual headwaters are often small and may not always exhibit high local richness, they collectively contribute disproportionately to regional (beta) diversity because many taxa are restricted to one or a few headwater catchments \citep{HeadwaterDiversity}. Because headwaters are the sources of all stream networks, their timing of connectivity to larger channels affect how quickly and broadly biotic signals, propagates, and community signatures spread downstream, and thus how local alterations (e.g., land use, forestry, pesticides) can scale up to affect network-wide diversity patterns \citep{Benda2004a,Fritz2018,HeadwaterDiversity}.

Therefore, as mentioned in \cite{Zaliapin2010} the development of a systematic framework which allow us to understand the processes on river networks capturing also the dynamical processes and not only the statical ones is of considerable theoretical and practical interest in several areas including hydrology and river ecology. In this work we follow the approach given in \cite{Zaliapin2010}, were one of such frameworks is constructed via the so called dynamic tree. The river network (a tree directed graph) or "static tree" encodes the topology of the drainage network, whereas the dynamic tree encodes the time-oriented downstream transport of the fluxes, hence, is constructed upon the static topology and at the same time integrates the dynamics of the fluxes.  This dynamic representation allow the authors to show a phase-transition behavior in how clusters of contributing sources coalesce as transport proceeds downstream, and provides a natural metric space in which to quantify distances between nodes based on their connectivity in time \citep{Zaliapin2010}, moreover this metric space is ultrametric. 

The dynamic-tree framework has recently been extended to identify ``dynamic clusters'' of sources whose fluxes merge at the same downstream junction under given topological conditions \cite{Roy2022}. In \cite{Roy2022}, the authors introduce two complementary connectivity criteria, structural extent and time of concentration, to characterize how side-branching controls the timing and rate at which subcatchments achieve full connectivity. This approach reveals why certain topological configurations either accelerate or delay the aggregation of fluxes toward the basin outlet.

In parallel, graph-theoretic analyses have been used to define critical nodes  in (static-) river networks, locations whose removal or failure most disrupts connectivity, travel times \cite{Sarker2019}.  Such work demonstrates that it is possible to translate concepts from complex networks into computational tools for identifying structurally important locations in real river basins.

Finally, network topology has been directly linked to the delivery of riverine ecosystem services.  Using synthetic networks with contrasting branching structures, \cite{Karki2021} showed that different topologies, for example, those in which many headwaters coalesce early versus those where sources remain separated over long distances, exhibit markedly different capacities for water supply, hydropower generation, sediment retention, nutrient uptake, flood attenuation, and aquatic habitat provision.  Networks with long, low-gradient main stems can attenuate floods and retain nitrate more effectively but may also be more susceptible to transmission losses, altering water-availability patterns \citep{Karki2021}.  These results reinforce earlier work showing that tributary position and confluence geometry govern physical heterogeneity and biological diversity at junctions \citep{Rice2006,Benda2004a} and suggest that managing connectivity and confluence structure is central to controlling downstream ecosystem functions.

Taken together, these studies show that (i) river networks can be naturally viewed as trees in which headwater sources connect to a common outlet; (ii) the timing and position at which sources join the main flow strongly influence hydrologic response, solute transport, and habitat structure; and (iii) graph- and tree-based measures can be used to identify critical or highly influential nodes.  Dynamic trees \citep{Zaliapin2010} and dynamic clusters \citep{Roy2022} offer a particularly appealing framework, because they encode not only who is connected to whom, but also \emph{when} that connectivity is realised along the network.  Within such a framework, it is natural to ask how to quantify the ``centrality'' of individual headwater sources in terms of their role in the time-oriented connectivity of the basin, for example, how early they join other sources, how many other sources they help integrate, and how strongly they influence downstream states.  Developing and analysing such centrality measures on the dynamic-tree (ultrametric) representation of a river network provides a way to bridge classic geomorphological theory, modern ecohydrological insights, and graph-theoretic notions of criticality into a single, coherent description of headwater importance.

\section{Dynamic trees for rivers}

We represent the drainage network by a rooted, directed tree
\(T_S = (V,E)\), where the root \(o \in V\) denotes the basin outlet
and the leaves \(X \subset V\) denote the sources of the river.
Each edge \(e \in E\) is oriented downstream and is assigned a
physical length \(l(e) > 0\), representing the distance between
two stream junctions along the river.

For any two nodes \(x,y \in V\) such that the unique directed path
from \(x\) to \(y\) is denoted by \(\gamma_y(x)\), we define the
length of this path as
\[
  L(x,y) := \sum_{e \in \gamma_y(x)} l(e).
\]

We assume that the travel time of the flow along an edge
\((u \to v)\) is a function of its length and the local velocity.
In the simplest case of constant velocity \(v(t) \equiv v_0 > 0\)
throughout the network, the travel time along a path
\(\gamma_y(x)\) is
\[
  \tau(x,y) := \frac{L(x,y)}{v_0}.
\]

For two sources \(a,b \in X\), let \(LCA_S(a,b) \in V\) denote
their least common ancestor in the static tree \(T_S\).
Following \cite{Zaliapin2010}, we define the (downstream)
\emph{meeting time} between \(a\) and \(b\) as
\begin{equation}\label{eq:meeting-time}
  d(a,b)
  := \max\bigl\{\tau\bigl(a,LCA_S(a,b)\bigr),
                \tau\bigl(b,LCA_S(a,b)\bigr)\bigr\}.
\end{equation}
then, \(d(a,b)\) is the first time
when their fluxes meet at their confluence.

We can construct an ultrametric space describing dynamic properties of the river network using the \emph{single-linkage hierarchical clustering} described as follows. Let \((X,d)\) be a finite metric space, with \(X=\{x_1,\dots,x_N\}\).
At the initial moment each element of \(X\) is regarded as an
individual cluster.  Thus we begin with the family of singleton
clusters
\[
  C_i^{(0)} := \{x_i\}, \qquad i=1,\dots,N,
\]
and denote by
\[
  \mathcal{P}^{(0)} := \bigl\{ C_1^{(0)},\dots,C_N^{(0)} \bigr\}
\]
the corresponding partition of \(X\).
Each cluster is represented in the tree by a leaf placed at height
\(t_0 := 0\).

Even at this initial stage the single–linkage dissimilarity between
two clusters is well-defined:
for distinct \(i,j\),
\[
  D^{(0)}(i,j)
  := \min\{ d(x,y) : x\in C_i^{(0)},\, y\in C_j^{(0)} \}
  = d(x_i,x_j),
\]
since the clusters are singletons.
Because \(X\) is finite there exists a pair of indices \((i_1,j_1)\)
attaining the global minimum
\[
  D^{(0)}(i_1,j_1)
  = \min_{1\le i<j\le N} d(x_i,x_j).
\]
We declare that the clusters \(C_{i_1}^{(0)}\) and \(C_{j_1}^{(0)}\)
merge at time
\[
  t_1 := D^{(0)}(i_1,j_1).
\]
Their union
\[
  C_\ast^{(1)} := C_{i_1}^{(0)} \cup C_{j_1}^{(0)}
\]
forms the first nontrivial cluster.  The partition at time \(t_1\) is
\[
  \mathcal{P}^{(1)}
  := \bigl( \mathcal{P}^{(0)}
             \setminus \{C_{i_1}^{(0)}, C_{j_1}^{(0)}\} \bigr)
     \cup \{ C_\ast^{(1)} \}.
\]
In the tree representation this merging introduces the first
internal vertex, placed at height \(t_1\), whose two children are
the leaves corresponding to \(x_{i_1}\) and \(x_{j_1}\).

Assume now that for some \(k\ge 1\) we have already constructed a
partition
\[
  \mathcal{P}^{(k)}
  = \{ C_1^{(k)},\dots,C_{m_k}^{(k)} \}
\]
of \(X\), together with merge times
\[
  0 = t_0 < t_1 < \dots < t_k,
\]
and a tree in which each cluster \(C_i^{(k)}\) corresponds to a
unique vertex lying at height \(\le t_k\) and consistent with all
previous merging events.

For any distinct clusters \(C_i^{(k)},C_j^{(k)}\) their single–linkage
separation is
\[
  D^{(k)}(i,j)
  := \min\{ d(x,y) : x\in C_i^{(k)},\ y\in C_j^{(k)} \}.
\]
Since the family \(\mathcal{P}^{(k)}\) is finite, there exists 
\((i_{k+1},j_{k+1})\) such that
\[
  D^{(k)}(i_{k+1},j_{k+1})
  = \min_{1\le i<j\le m_k} D^{(k)}(i,j).
\]
The clusters \(C_{i_{k+1}}^{(k)}\) and \(C_{j_{k+1}}^{(k)}\) merge at
time
\[
  t_{k+1} := D^{(k)}(i_{k+1},j_{k+1}),
\]
and their union
\[
  C_\ast^{(k+1)}
  := C_{i_{k+1}}^{(k)} \cup C_{j_{k+1}}^{(k)}
\]
forms the new cluster.  The next partition is
\[
  \mathcal{P}^{(k+1)}
  := (\mathcal{P}^{(k)} \setminus
      \{ C_{i_{k+1}}^{(k)}, C_{j_{k+1}}^{(k)} \})
     \cup \{ C_\ast^{(k+1)} \},
\]
and the tree is extended by creating a new internal vertex at height
\(t_{k+1}\) with children corresponding to the two merging clusters.

Each merging step reduces the number of clusters by exactly one, so
after \(N-1\) steps the partition consists of the single cluster
\(X\), whose associated vertex becomes the root of the tree.
The resulting structure is a rooted binary tree with leaves in
bijection with \(X\) and internal vertices labelled by the merge
times.  For any two leaves \(a,b\in X\) the height of their least
common ancestor equals the earliest time at which the clusters
containing \(a\) and \(b\) became identical in the above construction.
This height defines the induced ultrametric
\begin{equation} \label{eq:ultrametricrivermetric}
     d_u(a,b)
  := h\bigl( \operatorname{LCA}(a,b) \bigr).
\end{equation}

\begin{definition}
Let \((X,d)\) be the metric space attached to a  river network \(T_S\), where \(d\) is defined as in \ref{eq:meeting-time}. The dynamic tree \(T_D\) is the ultrametric tree representing the ultrametric space \((X,d_u)\) obtained by the single-linkage hierarchical clustering. 
\end{definition}

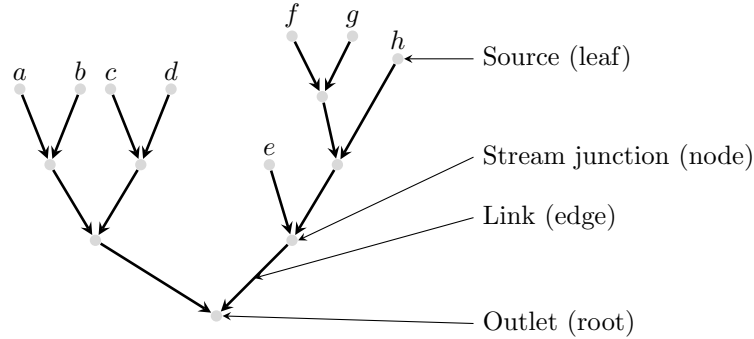
\begin{figure}[h!]
    \centering
\begin{tikzpicture}[scale=1,
    node/.style={circle,fill=gray!30,inner sep=1.5pt},
    treeedge/.style={line width=1pt,->,>=stealth},
    labarrow/.style={->,>=stealth}
]

\node[node] (r) at (0,0) {};

\node[node] (vL)  at (-1.6,1.0) {};
\node[node] (vAB) at (-2.2,2.0) {};
\node[node] (vCD) at (-1.0,2.0) {};
\node[node] (a)   at (-2.6,3.0) {};
\node[node] (b)   at (-1.8,3.0) {};
\node[node] (c)   at (-1.4,3.0) {};
\node[node] (d)   at (-0.6,3.0) {};

\node[node] (vR)   at (1.0,1.0) {};

\node[node] (e)    at (0.7,2.0) {};  
\node[node] (vRH)  at (1.6,2.0) {};

\node[node] (vFG)  at (1.4,2.9) {};
\node[node] (f)    at (1.0,3.7) {};
\node[node] (g)    at (1.8,3.7) {};
\node[node] (h)    at (2.4,3.4) {};  

\draw[treeedge] (vL) -- (r);
\draw[treeedge] (vR) -- (r);

\draw[treeedge] (vAB) -- (vL);
\draw[treeedge] (vCD) -- (vL);

\draw[treeedge] (a) -- (vAB);
\draw[treeedge] (b) -- (vAB);

\draw[treeedge] (c) -- (vCD);
\draw[treeedge] (d) -- (vCD);

\draw[treeedge] (e)   -- (vR);
\draw[treeedge] (vRH) -- (vR);

\draw[treeedge] (vFG) -- (vRH);
\draw[treeedge] (h)   -- (vRH);

\draw[treeedge] (f) -- (vFG);
\draw[treeedge] (g) -- (vFG);

\node[above] at (a) {$a$};
\node[above] at (b) {$b$};
\node[above] at (c) {$c$};
\node[above] at (d) {$d$};
\node[above] at (e) {$e$};
\node[above] at (f) {$f$};
\node[above] at (g) {$g$};
\node[above] at (h) {$h$};


\node[right] (labSource) at (3.4,3.4) {Source (leaf)};
\draw[labarrow] (labSource.west) -- (h.east);

\node[right] (labJunc) at (3.4,2.1) {Stream junction (node)};
\draw[labarrow] (labJunc.west) -- (vR.east);

\node[right] (labLink) at (3.4,1.3) {Link (edge)};
\coordinate (edgePoint) at (0.5,0.5);
\draw[labarrow] (labLink.west) -- (edgePoint);

\node[right] (labOutlet) at (3.4,-0.1) {Outlet (root)};
\draw[labarrow] (labOutlet.west) -- (r.east);

\end{tikzpicture}
    
    \caption{River network example. A network can be represented as a directed tree network, the leafs correspond to the sources, the internal nodes are the junctions, edges represent the links between those junctions and finally, the root represent the outlet of the river basin. }
    \end{figure}

One may interpret the ultrametric distance \(d_u(a,b)\)
through the following global coloring experiment. At time \(t=0\)
a dye is released simultaneously at \emph{all} sources in the basin.
The dye propagates downstream with constant velocity \(v_0\), and at
any time \(t>0\) we consider the portion of the river network that
has already been reached by the dye emitted from any source. This
yields a family of colored subgraphs that grow monotonically in time
and may consist of several disconnected components.

\begin{figure}[H]
    \centering
    \includegraphics[width=\linewidth]{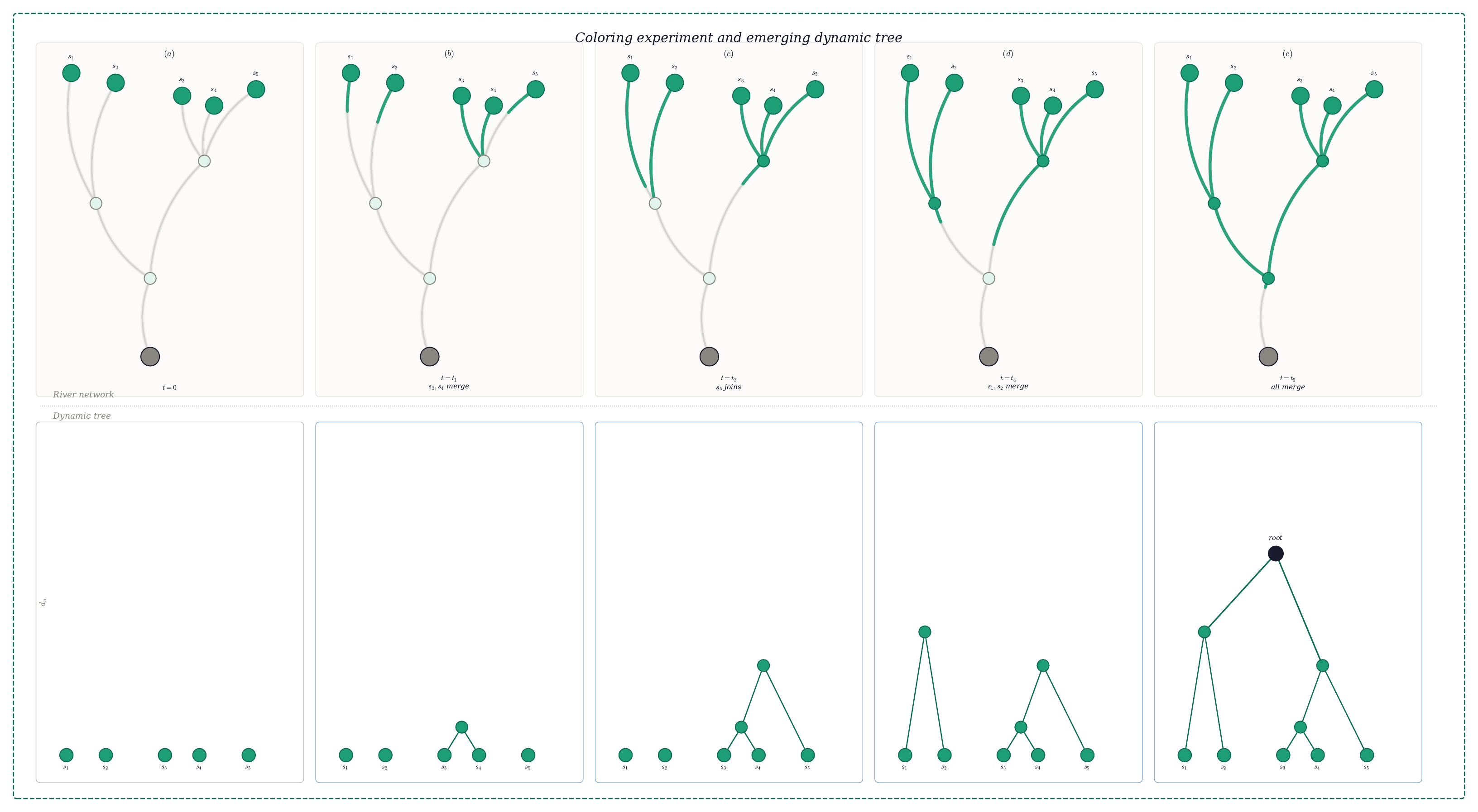}
    \caption{Coloring experiment illustrating the formation of dynamic 
    clusters (top) and the emerging dynamic tree (bottom). At $t=0$ a 
    dye is released simultaneously at all sources and propagates 
    downstream. Sources whose flows meet first form a dynamic cluster; 
    the height of each internal node in the dynamic tree records the 
    meeting time $d_u$ of the corresponding cluster.}
    \label{fig:coloring}
\end{figure}

Two sources \(a\) and \(b\) are said to belong to the same dynamic
cluster at time \(t\) if the colored subgraph contains a connected
path joining them; equivalently, if the fluxes originating from the
sets of sources that have already merged with \(a\) and with \(b\)
have become indistinguishable by time \(t\). The ultrametric
\(d_u(a,b)\) is the earliest time at which this occurs. In other
words, \(d_u(a,b)\) is the smallest time at which the evolving colored
network becomes connected between \(a\) and \(b\), so that their
dynamically formed sub-basins have fused into a single hydrologic
unit. This is schematized in Figure \ref{fig:coloring}. \newline

The next example is presented in \cite{Zaliapin2010}, will serve as a toy model to lustrate our analysis. The following image shows a sub basin of the Upper Noyo basin.   

\begin{figure}[H]\label{fig:dynamictreenoyo}
    \centering
    \includegraphics[width=\linewidth]{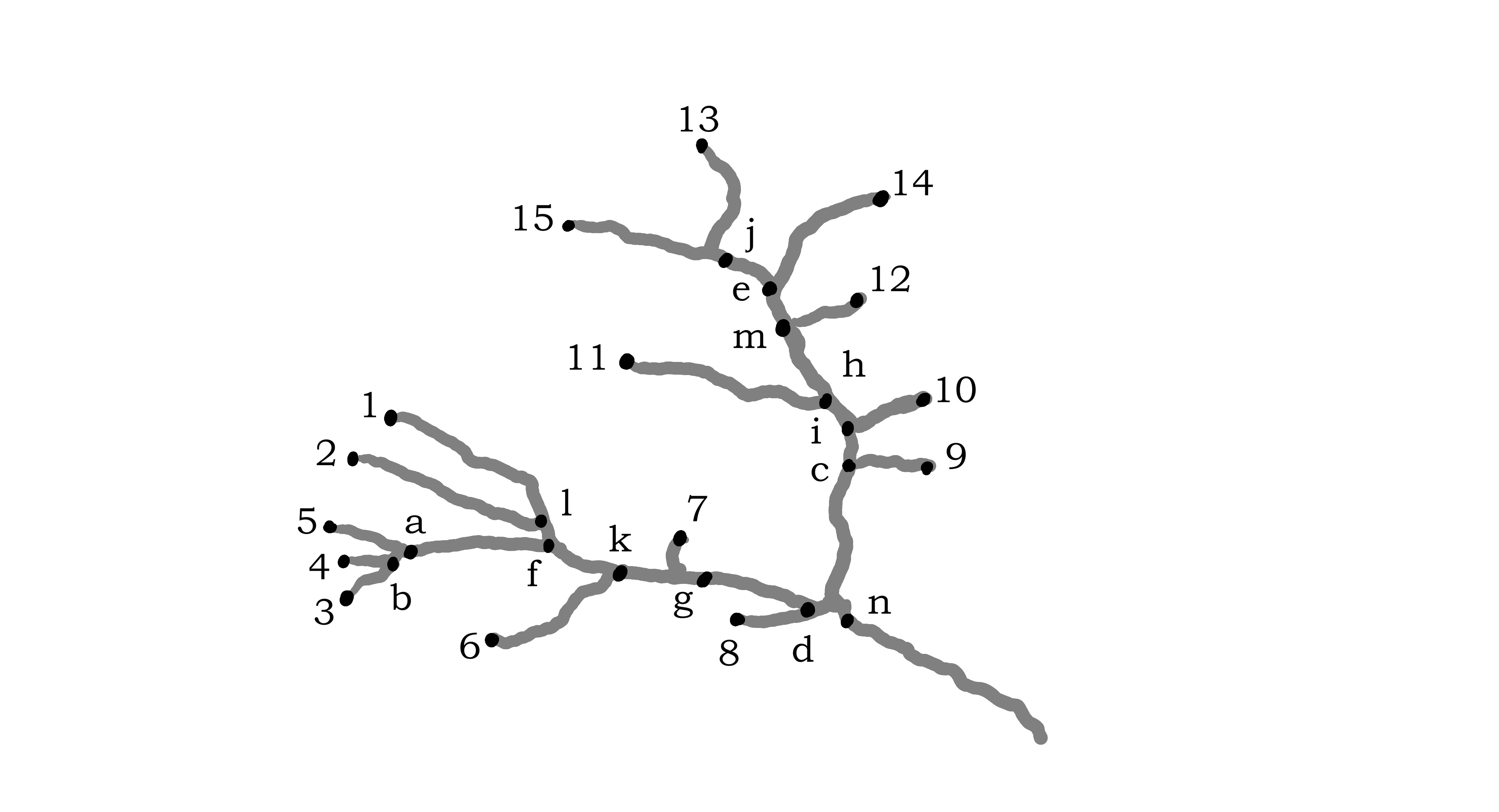}
    \caption{River network of a sub-basin of the Noyo river in Mendocino County, California. The sources are labeled by numbers while the stream junctions by letters. }
    \label{fignoyobasinriverreal}
\end{figure}
It is worth to mention some important distinctions between the dynamic tree and the river network. On one hand, the topological tree of the dynamic tree is not necessarily the same as the river network. For example, in the river network, the flux coming from the source \(13\) share with \(15\) the junction node \(j\). Topologically, "it seems" that \(15\) and \(13\) meet \(14\) "after" in node \(e\). Nevertheless, if the flux from source \(15\) meet first the source \(14\), due to velocity or distance reasons, in the dynamic tree the nodes \(15\) and \(13\) will be join after, the union of \(14\) and \(15\). 

\begin{figure}[H]
    \centering
    \textbf{(a)}\\[0.3em]
    \includegraphics[width=0.85\linewidth]{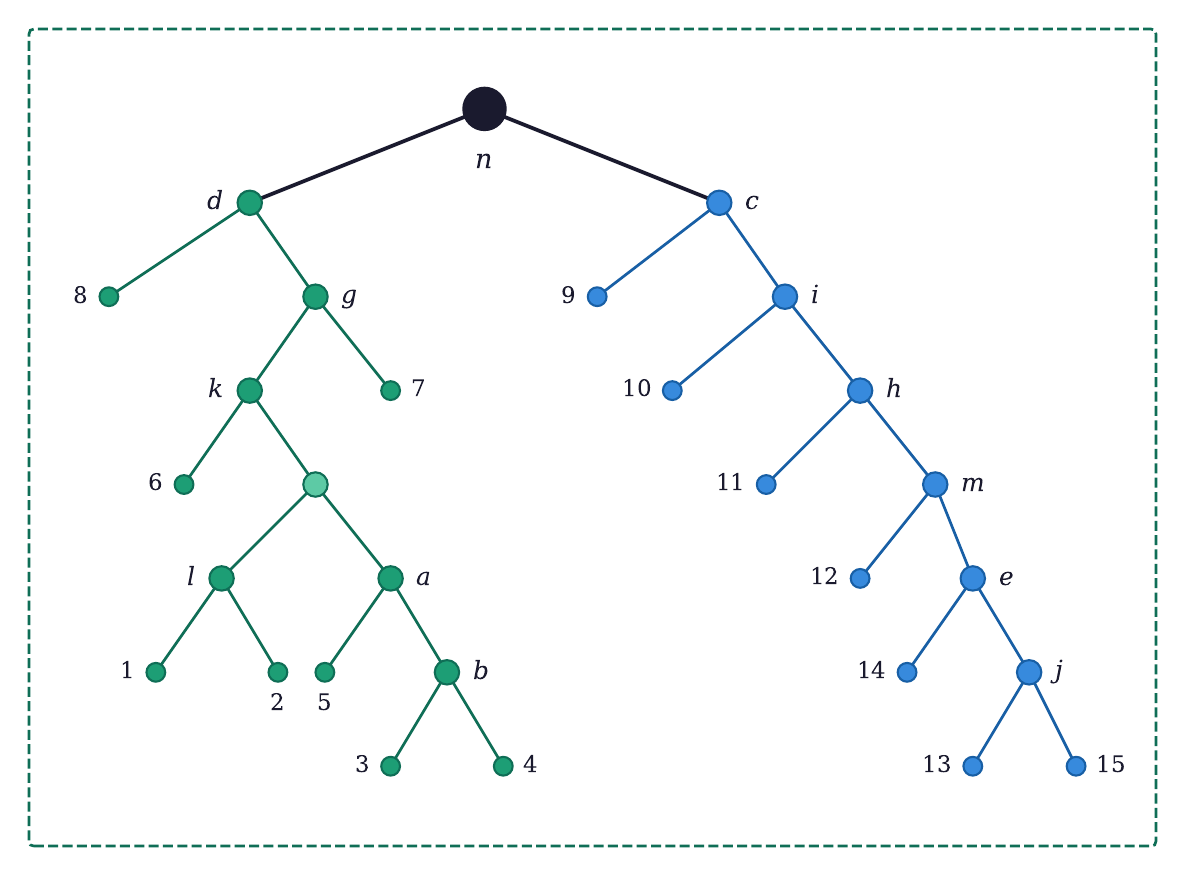}\\[1em]
    \textbf{(b)}\\[0.3em]
    \includegraphics[width=0.85\linewidth]{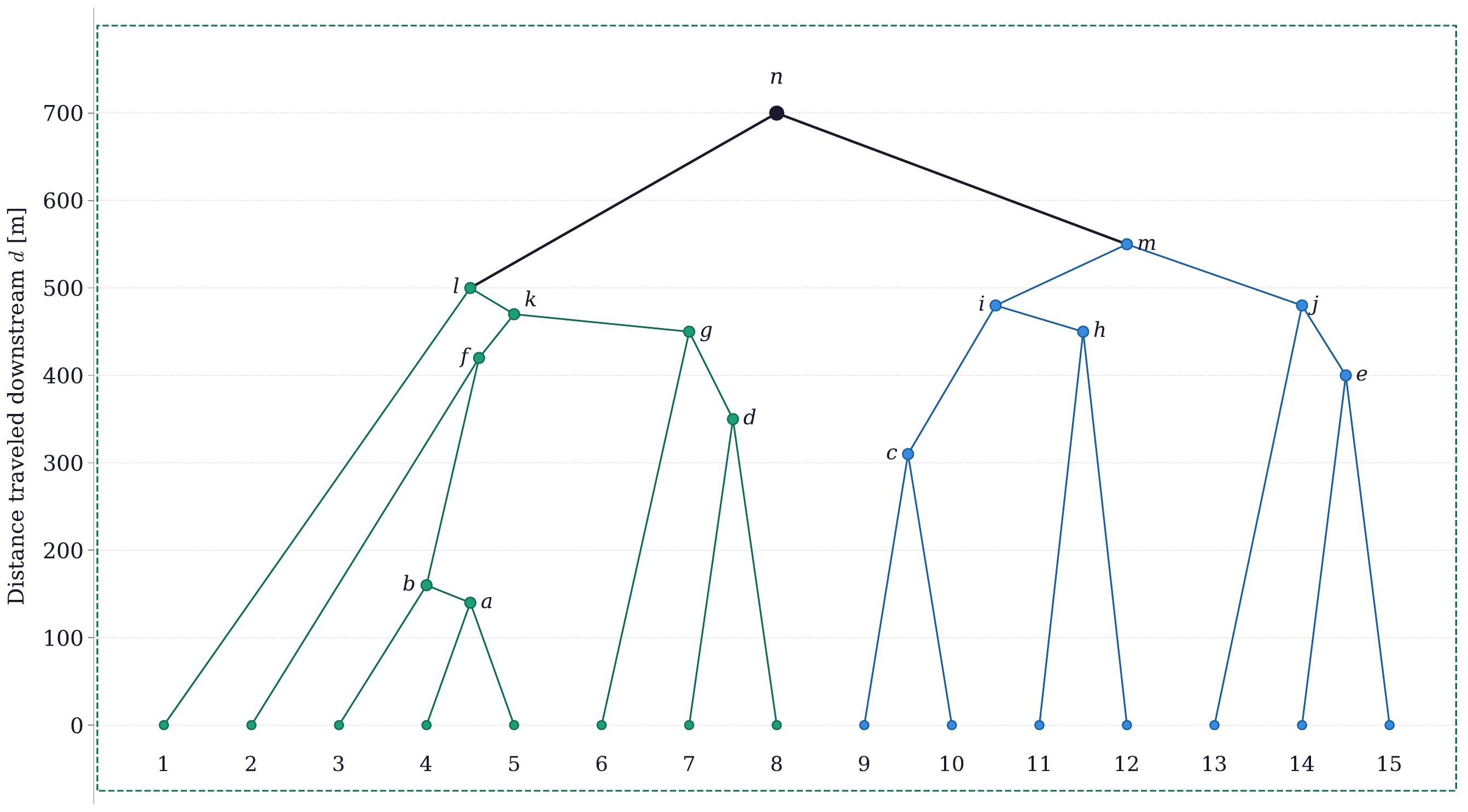}
    \caption{(a) The river network skeleton of the Noyo basin and 
    (b) the dynamic tree. Internal nodes represent stream junctions 
    and leaves represent headwater sources; the height of each 
    internal node in (b) corresponds to the meeting-time distance 
    $d_u$ between the sources in its subtree.}
    \label{fig:dynamictreenoyo}
\end{figure}
If we took the constant velocity to be \(v_0=1\), we can interpret \(d(a,b)\) as a distance between the sources as show in figure \ref{fig:dynamictreenoyo}. Another observation is worth to make. Note that node \(k\) in the dynamic tree has as leafs the nodes \(7\) and \(8\), where physically the fluxes coming from those sources never touch such junction. The meaning of \(k\) in the dynamic tree is the following: every internal node represent the junction where the fluxes from the sources in the respective sub tree finally meet; the dynamic tree describes the time-dependent history of the merging of the streams \cite{Zaliapin2010}. In this case is in \(k\) where all the streams from sources \(3-8\) meet for the first time assuming constant velocity.

\section{Ultrametric Laplacian and dynamic centrality}

The dynamic tree naturally gives rise to an ultrametric space via the 
meeting-time metric defined in equation~\eqref{eq:ultrametricrivermetric}: 
each source corresponds to a leaf, while the ultrametric structure encodes 
the dynamical and topological properties of the river network. Analyzing 
the geometry of a space through the heat kernel of an associated diffusion 
operator is a standard technique in spectral geometry and geometric data 
analysis, with growing applications in machine learning 
\cite{Coifman2006}. As shown in \cite{MoranLedezma2026}, any ultrametric 
tree, and in particular the dynamic tree of a river network, carries 
a canonical diffusion operator acting on its set of leaves, whose spectral 
properties encode the hierarchical merging structure of the basin. \newline

Given the dynamic tree $T_D$ of a river network, the set of sources
$X = \{x_1,\ldots,x_N\}$ equipped with the meeting-time distance $d_u$
forms a finite ultrametric space $(X, d_u, m)$, where $m$ is the
normalized counting measure $m(x) = 1/N$.

\begin{definition}
The \emph{ultrametric Laplacian} associated to $(X, d_u, m)$ is the
operator
\[
L_X u(x) = \sum_{y \in X} k(d_u(x,y))\bigl(u(y) - u(x)\bigr)\,m(y),
\]
where $k : [0,\infty) \to (0,\infty)$ is a non-increasing kernel
function. Its matrix representation $L = (L_{xy})_{x,y \in X}$ has
entries
\[
L_{xy} =
\begin{cases}
k(d_u(x,y))\,m(y) & \text{if } y \neq x,\\
-\displaystyle\sum_{z \neq x} k(d_u(x,z))\,m(z) & \text{if } y = x.
\end{cases}
\]
\end{definition}

The kernel $k$ determines the transition rates of the continuous-time Markov chain (CTMC) and therefore 
its temporal sensitivity with respect to the topology induced by $d_u$: 
a rapidly decaying $k$ assigns high transition rates to pairs of sources 
with small meeting-time distance, thereby emphasizing early confluences 
near the headwaters, while a slowly decaying $k$ distributes weight more 
uniformly across the hierarchy, giving relatively greater influence to 
distant downstream junctions.

\begin{thm}[\cite{MoranLedezma2026}]
Let $(X, d_u, m)$ be a finite measurable ultrametric space with
ultrametric Laplacian $L_X$ and $m$ the normalized counting measure.
Then $L_X$ generates an irreducible continuous-time Markov chain with transition function
\[
P_t(x,y) = H_t(x,y)\,m(y),
\]
where $H_t$ is the heat kernel of $L_X$, defined by
\[
(e^{tL_X}u)(x) = \sum_{y \in X} H_t(x,y)\,u(y)\,m(y).
\]
The chain is reversible and $m$ is its unique stationary distribution,
that is, $\lim_{t\to\infty}P_t(x,y) = m(y)$ for all $x,y\in X$.
\end{thm}

The CTMC models a random walker jumping between headwater sources 
with transition rates determined by their meeting times $d_u$: if 
$k$ is non-increasing, sources whose flows merge early communicate 
at higher rates, whereas transitions between sources that merge far 
downstream are comparatively rare. The process carries no direct 
physical meaning, it is a stochastic process whose jump structure 
is governed entirely by the ultrametric topology induced by $d_u$. 
In this way, the spectral properties of its generator encode the 
geometry of the basin: this is analogous to the classical setting 
of heat kernels on Riemannian manifolds, where the spectrum of the 
Laplace--Beltrami operator encodes the shape of the underlying space 
\cite{Coifman2006, Rosenberg1997}, but here the geometry is discrete 
and ultrametric rather than smooth. 

\begin{figure}[H]
    \centering
    \includegraphics[width=\linewidth]{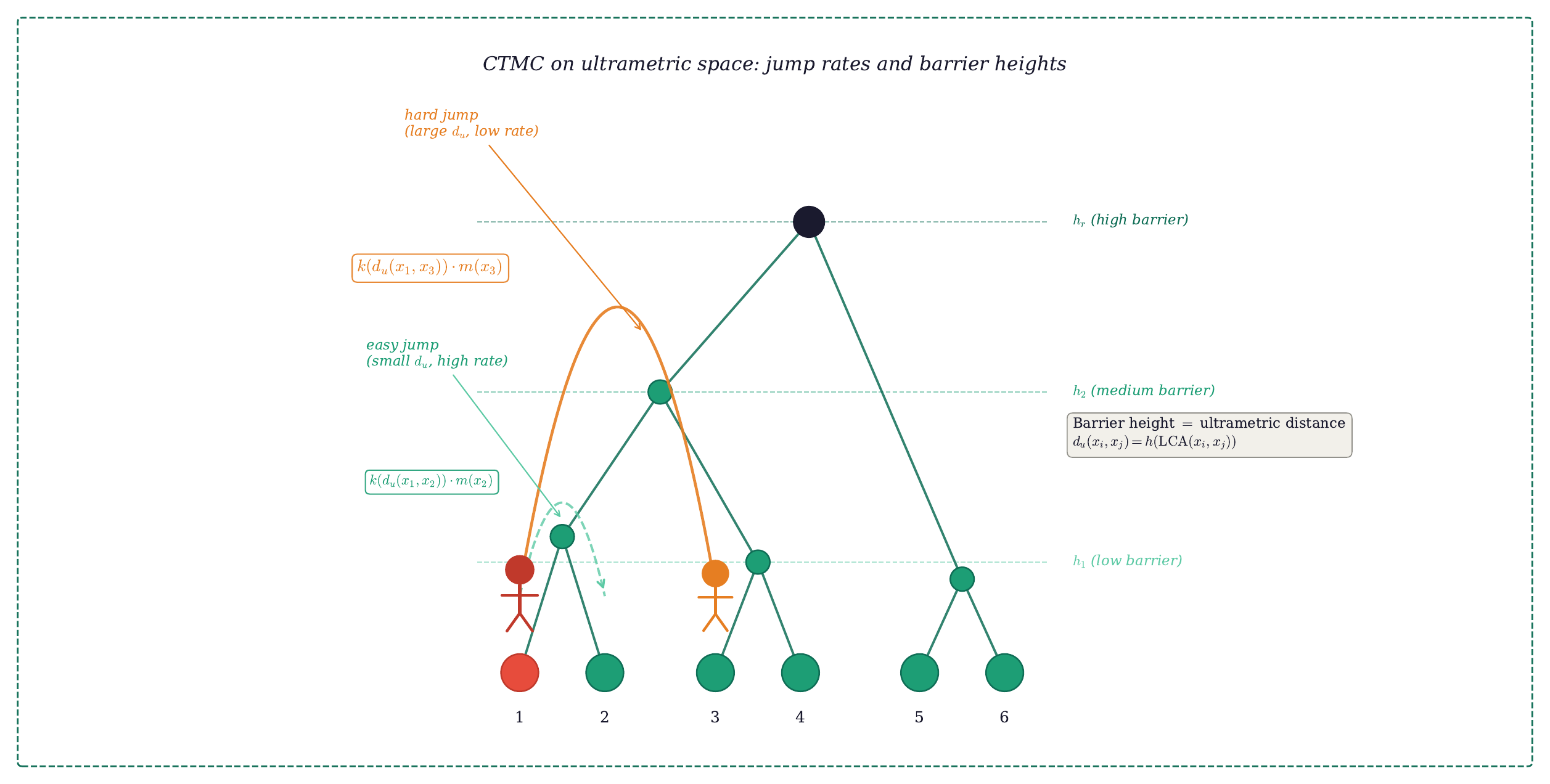}
    \caption{Schematic illustration of the CTMC on the ultrametric 
    space induced by the dynamic tree. A random walker jumps between 
    headwater sources with transition rates $k(d_u(x_i, x_j))\cdot m(x_j)$ 
    determined by their meeting-time distances. Short jumps (small $d_u$, 
    crossing low barriers) occur at high rates, while long jumps (large 
    $d_u$, crossing high barriers) are rare. The barrier heights correspond 
    to the ultrametric distances $d_u(x_i,x_j)=h(\mathrm{LCA}(x_i,x_j))$.}
    \label{fig:ctmc_walker}
\end{figure}

Each internal node $n \in T_D$ is associated with a ball 
$B_n \subset X$ consisting of all sources in its sub-tree, 
with diameter $\mathrm{diam}(B_n) = h(n)$, the height of 
$n$ in $T_D$.  A key structural fact 
is that the eigenvalues of $L_X$ are indexed by internal 
nodes: each node $n$ contributes exactly one eigenvalue 
$\lambda_n < 0$ with multiplicity $|C(n)|-1$, where $C(n)$ 
denotes the set of children of $n$. For a ball $T \subsetneq X$, 
we denote by $T^+$ its parent ball in $T_D$, and for a 
leaf $i \in X$, $F(i)$ denotes the parent node of $i$.

\begin{thm}[\cite{MoranLedezma2026}]\label{thm:heatkernel}
Let $(X,d_u,m)$ be a finite ultrametric space with dynamic tree $T_D$
and associated eigenvalues $\{\lambda_n\}$ indexed by internal nodes
$n \in T_D$. For any $x,y \in X$ with $LCA(x,y) = n$,
\begin{equation}\label{eq:heatkernel}
H_t(x,y) = 1 + e^{\lambda(B_n)t}
\left(\frac{\delta_{xy}}{m(\{x\})} - \frac{1}{m(B_n)}\right)
+ \sum_{\substack{T:\, B_n \subseteq T \subsetneq X}}
e^{\lambda(T^{+})t}
\left(\frac{1}{m(T)} - \frac{1}{m(T^{+})}\right),
\end{equation}
where the eigenvalues are given explicitly by
\begin{equation}\label{eq:eigenvalues}
\lambda_n = -\sum_{y \in X \setminus B_n} k(d_u(x_0,y))\,m(y)
- k(\mathrm{diam}(B_n))\,m(B_n), \quad x_0 \in B_n,
\end{equation}
and $T^{+}$ denotes the parent ball of $T$ in $T_D$.
\end{thm}

\begin{definition}[\cite{MoranLedezma2026}]\label{def:centrality}
For a source $i \in X$, the \emph{dynamic centrality index} is
\[
C_{\mathrm{CTMC}}(i)
= \left(\int_0^\infty \bigl(H_t(i,i) - 1\bigr)\,dt\right)^{-1}.
\]
Substituting \eqref{eq:heatkernel} into this expression yields the
closed-form formula
\begin{equation}\label{eq:centrality}
C_{\mathrm{CTMC}}(i) = \left(
-\frac{1}{\lambda(B_{F(i)})}
\left(\frac{1}{m(\{i\})}-\frac{1}{m(B_{F(i)})}\right)
+\sum_{\substack{T:\,B_{F(i)}\subseteq T \subsetneq X}}
-\frac{1}{\lambda(T^{+})}
\left(\frac{1}{m(T)}-\frac{1}{m(T^{+})}\right)
\right)^{-1},
\end{equation}
where $F(i)$ denotes the parent node of leaf $i$ in $T_D$.
\end{definition}

\begin{prop}[\cite{MoranLedezma2026}]\label{prop:mfpt}
For any two sources $i, j \in X$,
\[
C_{\mathrm{CTMC}}(i) > C_{\mathrm{CTMC}}(j)
\iff m_{ij} > m_{ji},
\]
where $m_{ij} = \mathbb{E}_i[T_j]$ is the mean first-passage time
from source $i$ to source $j$ under the CTMC generated by $L_X$.
\end{prop}

As a consequence of Proposition~\ref{prop:mfpt}, a source $i$ with 
higher dynamic centrality is easier to reach from any other source 
under the CTMC: $C_{\mathrm{CTMC}}(i) > C_{\mathrm{CTMC}}(j)$ if 
and only if $m_{ji} > m_{ij}$, meaning that the random walker 
arrives at $i$ from $j$ faster, on average, than it arrives at $j$ 
from $i$. When the underlying ultrametric space is the dynamic tree 
of a river network, this has a concrete topological reading: a 
headwater with high dynamic centrality is better connected to the 
rest of the sources in the dynamic sense; its flow integrates 
more rapidly into the network than that of a less central source. 
In this way, the CTMC does not model a physical process but rather 
acts as a spectral probe of the ultrametric topology, that reflects the dynamic connectivity structure of the basin.

\subsection{Topological interpretation}

The explicit formula~\eqref{eq:centrality} reveals how the topology
of the dynamic tree $T_D$ determines the centrality of each source
directly from its hierarchical structure. Two structural parameters
govern $C_{\mathrm{CTMC}}(i)$: the \emph{height} of the balls in
the ancestral path of $i$ toward the root, and the \emph{mass}
(number of leaves) contained in each such ball. Since both
quantities are read directly from $T_D$, no matrix diagonalization
or simulation of the stochastic process is required.

As a consequence, once the eigenvalues of $L_X$ are precomputed in
$O(n)$ time from the structure of $T_D$ \cite{MoranLedezma2026},
the centrality index $C_{\mathrm{CTMC}}(i)$ for all sources is
evaluated in $O(n)$ total time by a single depth-first traversal
of the tree. This makes the ultrametric spectral analysis scalable
to large river basins, in contrast to simulation-based approaches
whose cost grows with the number of time steps and sources.

\begin{prop}[\cite{MoranLedezma2026}]\label{prop:topology}
Let $k$ be non-increasing. Then:
\begin{enumerate}[label=\textup{(\roman*)}]
    \item \textup{(Height effect)} If the radius $\mathrm{diam}(B_n)$
    of a ball $B_n$ in the ancestral path $\gamma_{\mathrm{root}}(i)$
    decreases, then $C_{\mathrm{CTMC}}(i)$ increases. That is,
    sources whose flows merge earlier downstream have higher
    centrality.
    \item \textup{(Mass effect)} If the number of leaves $|B_n|$
    of a ball $B_n$ in $\gamma_{\mathrm{root}}(i)$ increases, then
    $C_{\mathrm{CTMC}}(i)$ increases. That is, sources belonging
    to larger dynamic clusters are more central.
    \item \textup{(Level-regular case)} If all internal nodes of
    $T_D$ at the same depth have the same number of children and
    the same height, then $C_{\mathrm{CTMC}}(i) = C_{\mathrm{CTMC}}(j)$
    for all $i,j \in X$.
\end{enumerate}
\end{prop}

\begin{figure}[h!]
\centering
    \includegraphics[width=\linewidth]{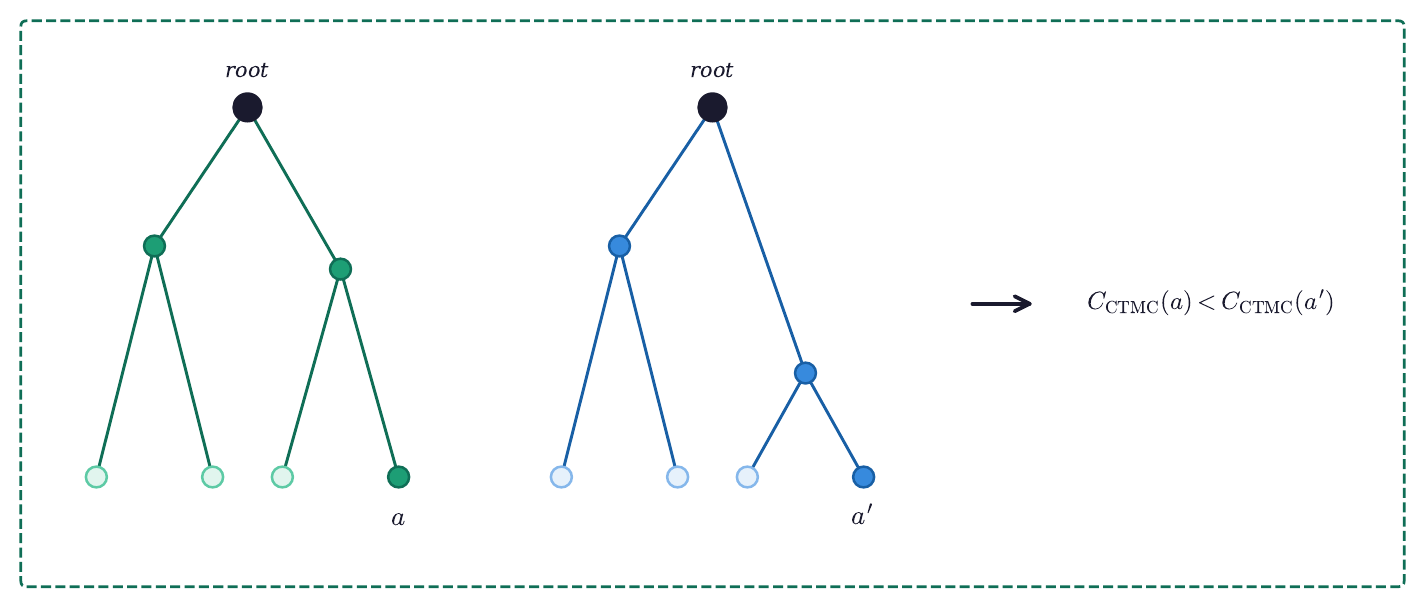}
\caption{Height effect: decreasing the merger height of the ball
containing $a'$ increases its dynamic centrality relative to $a$.
In river terms, a headwater whose flow merges earlier with its
neighbours is more central than one that remains isolated longer.}
\end{figure}

\begin{figure}[h!]
\centering
    \includegraphics[width=\linewidth]{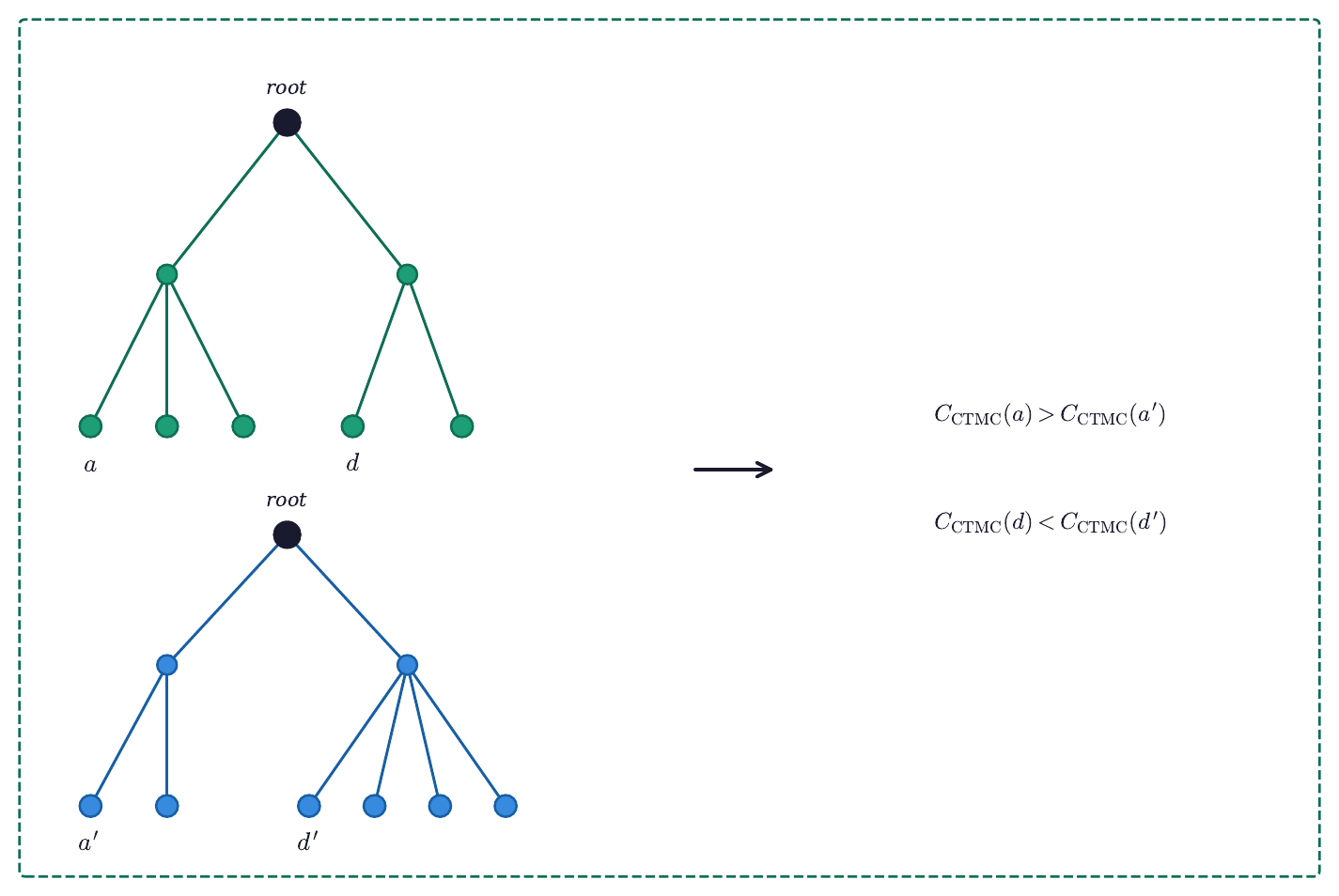}
\caption{Mass effect: increasing the number of leaves of the ball
containing $d'$ increases its dynamic centrality relative to $d$,
while reducing leaves in the ball of $a'$ decreases its centrality
relative to $a$. In river terms, a headwater belonging to a larger
dynamic cluster is more accessible to the random walk.}
\end{figure}

To make these structural properties concrete, consider the Noyo 
basin example of Figure~\ref{fig:dynamictreenoyo}. Sources $1$ 
and $4$ appear similarly connected in the static river network 
skeleton, yet their dynamic centralities differ markedly: by the 
time the flow from source $4$ reaches junction $k$, the flow from 
source $1$ has not yet arrived at node $l$, and when it finally 
does, the flows from sources $2$--$8$ have already merged. The 
path $4 \to n$ is more ramified in the dynamic tree, it passes 
through balls of higher mass at smaller heights, and therefore 
$C_{\mathrm{CTMC}}(4) > C_{\mathrm{CTMC}}(1)$, consistent with 
Proposition~\ref{prop:topology}. \textit{High-centrality 
headwaters are thus the tributaries that most efficiently 
transmit water into the rest of the network.} The empirical 
validation of this interpretation on a dataset of 49 natural 
river basins is presented in the next section.
\section{Centrality and reached junctions}

We use the dataset introduced by Roy et al.~\cite{Roy2022}, consisting of 49 natural river basins with minimal human impact, located across different geographic regions in the United States. The basins correspond to drainage areas ranging from 0.15 to 4.37\,km$^2$, and their channel networks were originally extracted from 1\,m resolution DEMs. \newline

\begin{figure}[H]
\centering
\includegraphics[width=\textwidth]{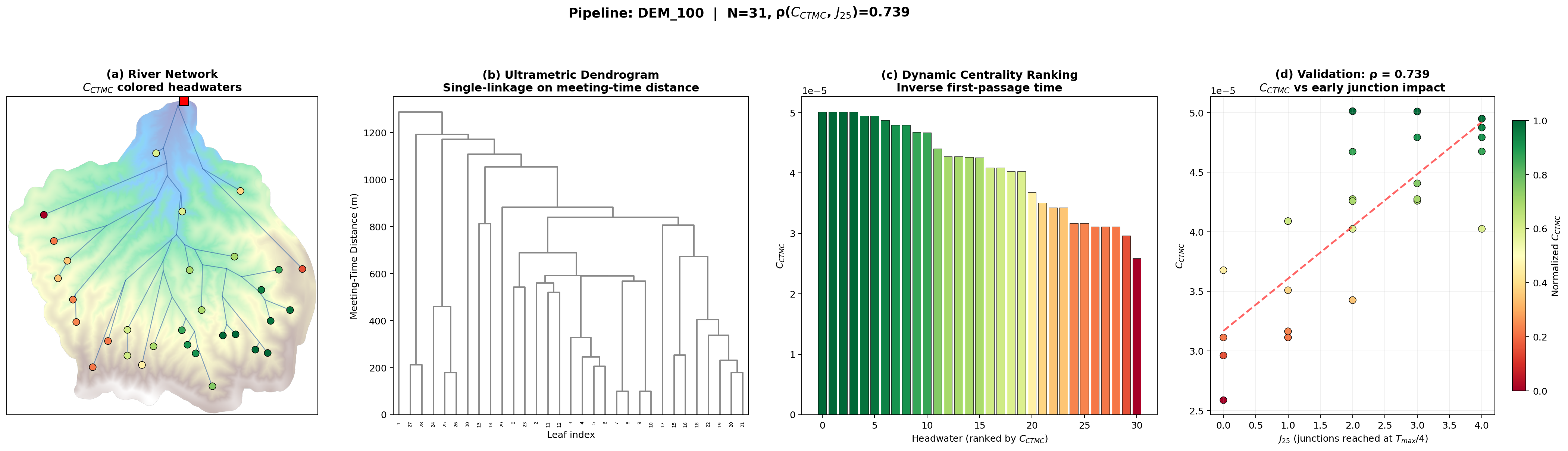}
\caption{Overview of the $C_{\text{CTMC}}$ pipeline applied to basin DEM\_100 ($N = 31$ headwaters). (a) Extracted river network ; headwaters are colored by their $C_{\text{CTMC}}$ value (green = high, red = low), and the outlet is marked by a red square. (b) Ultrametric dendrogram obtained via single-linkage clustering on the meeting-time distance matrix $M$, where the height at which two leaves merge corresponds to the time required for their respective fluxes to meet downstream. (c) Dynamic centrality ranking: headwaters sorted by decreasing $C_{\text{CTMC}}$, defined as the inverse of the mean first-passage time on the CTMC. (d) Validation: Spearman rank correlation between $C_{\text{CTMC}}$ and $J_{0.25}$, the number of junctions reached at $t = 0.25\, T_{\max}$. The positive correlation ($\rho = 0.739$) indicates that headwaters with higher dynamic centrality reach more junctions during the early phase of transport. }
\label{fig:pipelineultra}
\end{figure}

In order to assess the practical reliability of the centrality measure $C_{\text{CTMC}}$, we analyze how the top-ranked headwaters perform with respect to the number of junctions reached during the early phase of transport. For each basin, we select the top 25\% of headwaters according to $C_{\text{CTMC}}$ and determine whether they land in the upper half of the $J_{t}$ ranking, where $J_{t}$ denotes the number of junctions reached at transport time $t$, expressed as a fraction of the time at which all sources 
have reached the basin outlet denoted by $T_{\max}$. We repeat this procedure for the null model given by depth, that is, the distance from each headwater to the outlet. The pipeline is displayed in Figure \ref{fig:pipelineultra}.

It is worth emphasizing the conceptual distinction between 
$C_{\mathrm{CTMC}}$ and the junction-reach metric $J_t$ 
used for validation. $J_t$ is an empirical quantity that 
requires explicit simulation of the transport process: 
for each basin and each time $t$, one must propagate 
fluxes downstream and count the number of junctions 
reached by each source, incurring a computational cost 
that grows with both the number of sources and the number 
of time steps evaluated. In contrast, $C_{\mathrm{CTMC}}$ 
is computed \emph{a priori} from the ultrametric structure 
of the dynamic tree alone, without any simulation, in 
$O(n)$ time. It does not approximate $J_t$; rather, it 
predicts which headwaters will be dynamically influential 
before the transport is run. The strong agreement between $C_{\mathrm{CTMC}}$ rankings 
and $J_t$ performance across 45 out of 49 basins suggests 
that the spectral geometry of the ultrametric space 
captures relevant information about the physical transport 
dynamics of the basin. This is consistent with the 
probabilistic interpretation of $C_{\mathrm{CTMC}}$ via 
mean first-passage times (Proposition~\ref{prop:mfpt}), 
which grounds the index in the same hierarchical merging 
structure that governs downstream transport.

\begin{figure}[H]
    \centering
    \includegraphics[width=1\linewidth]{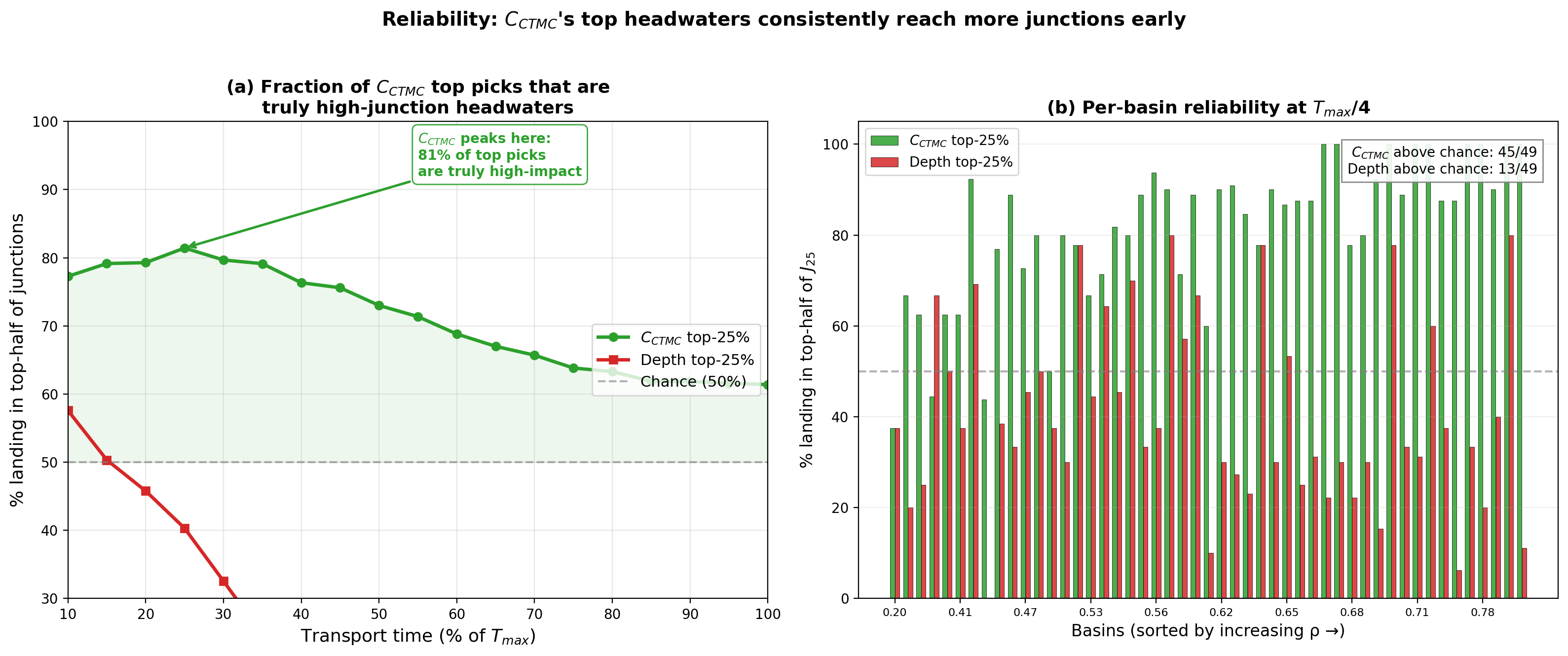}
    \caption{(a) Fraction of the CTMC ranking top picks that are truly high-junction headwaters. (b) Per-basin reliability at a quarter of total time.}
    \label{fig:reliability}
\end{figure}

Figure~\ref{fig:reliability}(a) shows the fraction of these top picks that are truly high-junction headwaters, as a function of transport time. We observe that during the early transport phase ($t < 0.4\, T_{\max}$), approximately 81\% of the headwaters identified by $C_{\text{CTMC}}$ are indeed among the high-junction headwaters, compared to 50\% expected by chance. In contrast, the depth-based selection drops below 40\% in this regime, performing worse than a random selection. Notice that the $C_{\text{CTMC}}$ curve remains above the chance level for all transport times, indicating that the headwaters identified as central are not only early contributors but consistently relevant throughout the entire transport process.

Figure~\ref{fig:reliability}(b) displays the per-basin reliability at $t = 0.25\, T_{\max}$: for each of the 49 basins, sorted by increasing Spearman rank correlation $\rho(C_{\text{CTMC}}, J_{0.25})$, we show the fraction of top 25\% headwaters that land in the upper half of the $J_{0.25}$ ranking. Here, $\rho$ measures the monotonic agreement between the full ranking given by $C_{\text{CTMC}}$ and the one given by $J_{0.25}$. We can observe that $C_{\text{CTMC}}$ surpasses the chance level in 45 out of 49 basins, while depth achieves this in only 13. 

\begin{figure}[H]
\centering
\includegraphics[width=\textwidth]{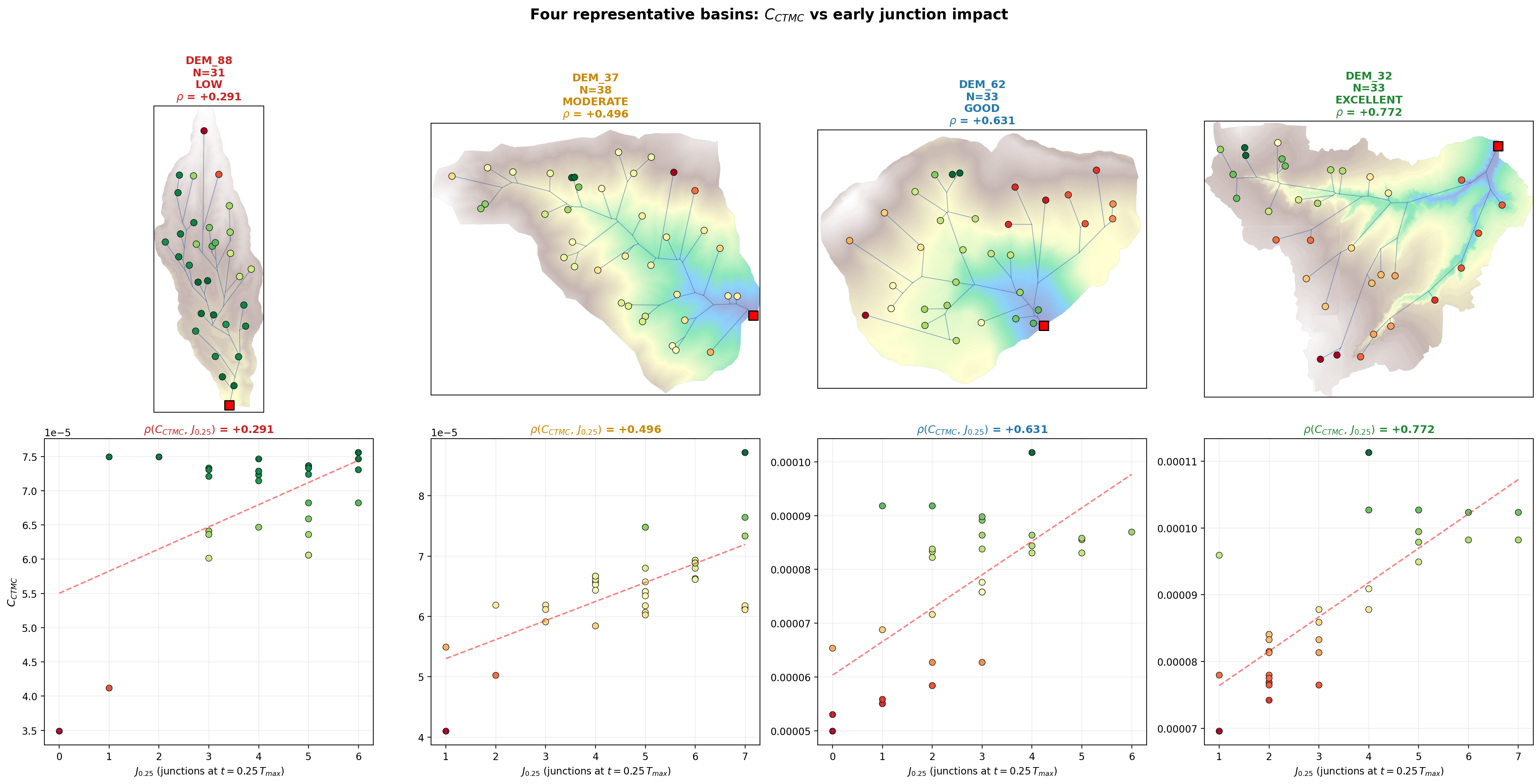}
\caption{Four representative basins illustrating different levels of agreement between $C_{\text{CTMC}}$ and early junction impact. Top row: extracted river networks overlaid on their respective DEMs, with headwaters colored by $C_{\text{CTMC}}$ (green = high, red = low). Bottom row: scatter plots of $C_{\text{CTMC}}$ against $J_{0.25}$, the number of junctions reached at $t = 0.25\, T_{\max}$. From left to right, the basins are ordered by increasing Spearman rank correlation $\rho$. Despite similar network sizes, the correlation ranges from low ($\rho = 0.29$) to excellent ($\rho = 0.77$), indicating that the predictive power of $C_{\text{CTMC}}$ depends on the spatial configuration of the network rather than on its size.}
\label{fig:tiers}
\end{figure}

In order to illustrate how the agreement between $C_{\text{CTMC}}$ and the early junction impact varies across different network topologies, we present in Figure~\ref{fig:tiers} four representative basins spanning the observed range of $\rho(C_{\text{CTMC}}, J_{0.25})$. We can observe that even in the basin with the lowest agreement (DEM\_88, $\rho = 0.29$), the general trend is positive: headwaters with higher $C_{\text{CTMC}}$ tend to reach more junctions early. As the correlation increases, this trend becomes progressively sharper, with the best case (DEM\_32, $\rho = 0.77$) showing a clear monotonic relationship. It is worth to mention that the four basins have similar numbers of headwaters ($N$ between 31 and 38),  but different correlations. This suggests that the performance of $C_{\text{CTMC}}$ depends not on the size of the network but on the specific spatial arrangement of its branches, an observation that is consistent across the full dataset of 49 basins.

\section{Conclusions}

We have justified and studied the development of a 
centrality index for headwaters in a river network. 
The centrality theory developed in the previous 
sections, which applies to CTMCs, is proposed as a 
natural tool for detecting headwater centrality using 
the dynamic tree framework of \cite{Zaliapin2010}, 
which introduces a way of analyzing the dynamics and 
topology of a river basin through an ultrametric space. 
By analyzing real data, we have found that this index 
carries a clear physical interpretation: it ranks higher 
those headwaters that connect most rapidly to the basin, 
and our study shows that in the majority of cases the 
highest-ranked headwaters are also those that pass 
through the most junctions along the path of the flow 
toward the outlet. The systematic detection of such 
headwaters could be an important tool in ecological 
care and prevention tasks, since the role of junctions 
and headwaters is key to the development of ecological 
mechanisms such as those mentioned in the introduction.

\begin{figure}[h!]
\centering
\includegraphics[width=\textwidth]{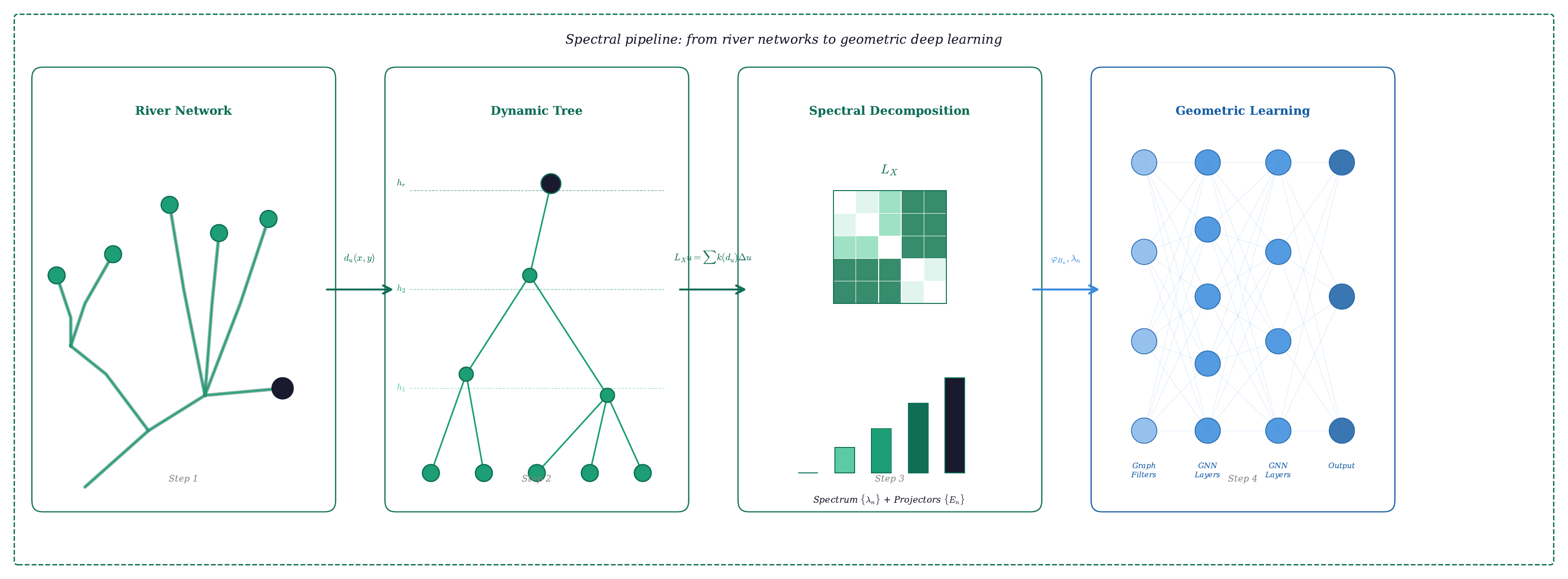}
\caption{Spectral pipeline: from river network to 
geometric deep learning. The ultrametric Laplacian 
$L_X$ provides the spectral basis for graph filters 
that feed graph neural network layers, enabling 
data-driven prediction of downstream transport from 
headwater measurements.}
\label{fig:pipeline}
\end{figure}

The explicit spectral decomposition of $L_X$ into 
eigenvalues $\lambda_n$ and eigenprojectors $E_n$ (see \cite{MoranLedezma2026}) 
suggests a natural extension toward graph signal 
processing and geometric deep learning. The 
eigenfunctions of $L_X$ provide a canonical spectral 
basis for defining graph filters adapted to the 
ultrametric geometry of the basin, analogous to 
the Fourier basis on Euclidean domains. This suggests a natural extension toward a learning-oriented pipeline in which 
field measurements at headwater sources such as 
nutrient concentrations, biodiversity indices, or 
sediment loads are filtered in the spectral domain 
of $L_X$ and fed into graph neural network layers 
to predict downstream transport or classify 
hydrologically critical sources. Figure~\ref{fig:pipeline} 
illustrates this pipeline schematically.

\section*{Data Availability}
The river basin dataset used in this study was introduced 
by Roy et al.~\cite{Roy2022} and consists of 49 natural 
river basins across the United States, originally extracted 
from 1\,m resolution digital elevation models (DEMs). 
The dataset is publicly available through the repository 
associated with \cite{Roy2022}. No new data were generated 
in this work.

\section*{Acknowledgements}

 Patrick Bradley is warmly thanked for the many useful conversations we had and for his general advice. This work is supported by the Deutsche Forschungsgemeinschaft
under project number 469999674.

\bibliography{biblio}

\end{document}